\documentclass[12pt]{article}

\newcommand{\ks}[1]{#1 \!\!\! \slash } 
\def\Title#1{\begin{center} {\Large {\bf #1} } \end{center}}

\begin{document}

\Title{Probing non-perturbative QCD through  hadronic matrix elements extracted from exclusive hard processes } 

\bigskip\bigskip


\begin{raggedright}  

{\it B.Pire$^1$\index{Pire,B} and L. Szymanowski$^2$\index{Szymanowski,L}\\
$1.$ CPHT - \'Ecole Polytechnique, CNRS,
F-91128 Palaiseau, FRANCE\\
$2.$ Soltan Institute for Nuclear Studies,
PL-00681 Warsaw, POLAND}
\bigskip\bigskip
\end{raggedright}

\section{Introduction}

QCD is the theory of strong interactions and non-perturbative methods have been developed to address the confinement property of QCD.
Many experimental measurements probe the confining dynamics, and it is well-known that hard scattering processes allow the extraction of non perturbative hadronic matrix elements. To study exclusive hard processes, such as electromagnetic form factors and reactions like $\gamma^* N \to \gamma N'$ , $\gamma^* N \to \pi N'$,  $\gamma^* \gamma \to \pi \pi$ , $\bar p p \to \gamma^* \pi$ in particular kinematics (named as generalized Bjorken regime), one introduces specific non-perturbative objects,  namely generalized parton distributions (GPDs), distribution amplitudes (DA)  and transition distribution amplitudes (TDA), which are Fourier transformed non-diagonal matrix elements of non-local operators $\bar \psi_\alpha(z)\psi_\beta(0) $ or $ \psi_\alpha(z)\psi_\beta(z')\psi_\gamma(0)$ on the light-cone. We review here a selected sample of exclusive amplitudes in which the quark and gluon content of hadrons is probed, and emphasize that much remains to be done to successfully compute their non-perturbative parts. We  present some difficulties with respect to the application of the much publicized AdS-QCD approach to the calculation of these partonic quantities.

\section{Accessing hadronic  Distribution Amplitudes}
\subsection{Nucleon and $\pi$ meson}
The basic properties of meson (and baryon) distribution amplitudes have been worked out in many details since the pioneering works \cite{DA}.
Conventionally, the three proton  DAs of twist 3 are defined through the decomposition of the 
 matrix element of the 3-quark operator in Eq.1
 in terms of three invariant functions of the scalar product of
the light-like separation $\tilde z_i$ with the proton 
momentum $p_p$, $V(\tilde z_i.p_p)$,$A(\tilde z_i.p_p)$ and $T(\tilde z_i.p_p)$,  
\begin{eqnarray}
&&\langle 0|u_\alpha(z_1 n)u_\beta(z_2 n)d_\gamma(z_3 n)|p_p)\rangle=\frac{1}{4}f_N \times 
 \Big[
V^p(\tilde z_i.p_p) ( \ks p_p C)_{\alpha \beta} (\gamma^5 u^+(p_p,s_p))_\gamma \nonumber\\
&& + ~~A^p(\tilde z_i.p_p) (\ks p_p \gamma^5 C)_{\alpha \beta} u^+(p_p,s_p)_\gamma
+T^p(\tilde z_i.p_p) (\sigma_{p_p  \mu}\,C)_{\alpha \beta} 
(\gamma^\mu \gamma^5 u^+(p_p,s_p))_\gamma
\Big] \;. 
\label{eq:DA}
\end{eqnarray}
The latter functions satisfy $V^p(\tilde z_i.p_p=0)=T^p(\tilde z_i.p_p=0)=1$ and  $
A^p(\tilde z_i.p_p=0)=0$,
which provides the interpretation of $f_N$ as the value of the proton wave function at the origin.
To go to  momentum space one writes a  Fourier transform 
which  enables to define functions of momentum fractions $x_i$. 

The asymptotic solutions derived from evolution equations do not appear to be phenomenologically adequate and inclusion of higher order Gegenbauer  (or Appel) polynomials is required. Various non-perturbative techniques have been used to quantify this statement, from lattice computations to QCD sum rules techniques. A recent careful comparison \cite{Lenz} of their results for the proton looks quite instructive since it shows large discrepancies between the models. For instance, the first moments estimated to (0.56, 0.19, 0.23) from QCD sum-rules are evaluated on the lattice as (0.40,0.30,0.30).  This is not so surprising given the different assumptions of each method. What is more worrying is that the expected (and quoted) theoretical uncertainties are notoriously inconsistent with the differences between the results. Some more humble way to estimate the theoretical uncertainties is obviously required.

Although the debate on the range of applicability of factorized amplitudes is still open, the scaling region seemed to be reached in the mesonic case for $Q^2 \approx$ a few $GeV^2$, when recent experimental data \cite{Babar} on the $\pi^0 \gamma$ transition form factors has  brought surprises. Quite new ideas \cite{shape} have been put forward to explain the data, which may look a little bit prematury given the uniqueness of the experimental analysis. A check of the strange $Q^2$ behaviour seen by the BABAR collaboration should be possible with the analysis of the two meson channel $\gamma^* \gamma \to \pi^0 \pi^0$ in the framework of the generalized $\pi^0 \pi^0$ distribution amplitudes \cite{GDA}.

\subsection{Hybrid meson}

Other exotic objects which can be investigated in hard reactions are the mesons with exotic quantum numbers, for example, with $ J^{PC} = 1^{-+}$. These
hybrid mesons cannot be described within the usual quark model. It has nonetheless been shown \cite{hybrid} that the fact that the quark-antiquark
correlator on the light cone includes a gluonic component due to gauge invariance, implies the existence of a leading
twist hybrid light-cone distribution amplitude. 
Using the results of seminal studies  on the DAs of vector mesons \cite{BB}, one writes ($\bar u =1-u$): 
\begin{eqnarray}
\label{hmeV}
&&\langle H(p,\lambda) |\bar\psi(z)\gamma_\mu [z;-z]\psi(-z)|0\rangle =
\nonumber\\
&&f_H M_{H} \biggl[
p_\mu e^{(\lambda)}\cdot n
\int\limits_{0}^{1} du\, e^{i(u-\bar u) p\cdot z} \phi_1^{H}(u)
+
e_{\mu\,T}^{(\lambda)} \int\limits_{0}^{1} du\, e^{i(u-\bar u) p\cdot z} \phi_3^{H}(u)
\biggr] \nonumber
\end{eqnarray}
for the vector correlator, and a similar equation for  the axial correlator. 
The polarization vector $e^{(\lambda)}_\mu$ describes the spin state 
of the hybrid meson.
Due to C-charge invariance, the symmetry properties of DAs are manifested in the 
fact that
$\phi_1^{H}(u)=-\phi_1^{H}(1-u), \quad \phi_3^{H}(u)=-\phi_3^{H}(1-u), \quad \phi_A^{H}(u)=\phi_A^{H}(1-u)$.
Compared to the $\rho$ meson matrix elements, one can see that 
the  corresponding DAs for the exotic hybrid meson 
have different symmetry properties. 

The leading twist longitudinally polarized 
hybrid meson DA is asymptotically equal to 
\begin{eqnarray}
\label{approxH}
\phi_{1}^{H}(u)=30 u (1-u)(1-2u) \nonumber
\end{eqnarray}
A non-perturbative estimate of the normalization $f_{H}$ \cite{fH} based on QCD sum rules techniques yields $f_H \approx 50 $ MeV. No lattice determination has to our knowledge been published.
Deep exclusive electroproduction processes should thus produce normal and hybrid mesons on the same footing (as far as the $Q^2$ behaviour is concerned). An experimental confirmation of this result is waited for. 

On the other hand, such a reasoning does not apply to isotensor tetraquark states, which have genuine twist 4 DAs. This allows  to have a way to distinguish them from $q \bar q$ excitations \cite{tetra}, provided there are produced in a hard process.

\subsection{Transversally polarized vector meson}
The distribution amplitudes of transversally polarized vector mesons needs special care. Indeed the leading twist DA for such a hadron is chiral-odd, and hence decouples from the most simple
hard amplitudes \cite{DGP} so that, to extract them from experimental data, one needs to study reactions with more than two final
hadrons \cite{IPST} or face difficulties connected with factorization breaking terms.

One thus has to deal with
the twist 3 more intricate part of the amplitude \cite{t3}.
It also requires the introduction of
 matrix elements of quark-antiquark-transverse gluon nonlocal operators
$ \langle \rho(p)|\bar\psi(z_1)\gamma_{\mu}g A_{\alpha}^T(z_2) \psi(0) |0\rangle$ and $\langle \rho(p)|
\bar\psi(z_1)\gamma_5\gamma_{\mu} g A_{\alpha}^T(z_2) \psi(0) |0\rangle$, the Fourier transform of which are parametrized by new DAs.

An understanding of the quark-gluon structure of a
transversally polarized vector meson is however an important task of hadronic physics. These twist 3 DAs have been analyzed in detail in \cite{t3DA} but lattice techniques have  not yet been applied  to their study.
This is a rich domain as far as  phenomenological studies are concerned, since many data on the electroproduction of vector mesons at various energies are and will be available. 

\section{Accessing the photon  Distribution Amplitude}
The photon is a very interesting object for QCD studies since it has both
a pointlike coupling to quarks, which yields a perturbative part of photonÕs wave function, and a 
non-perturbative coupling related to  the magnetic 
susceptibility of the QCD vacuum and which builds its
chiral-odd twist-2 distribution amplitude. This allows us 
to use a transverse spin asymmetry to  probe the chiral odd 
distribution amplitude of the photon.
In Ref.\cite{PRL}, we describe a new way to access the photon distribution amplitude through the  photoproduction of lepton pairs on a transversally polarized proton. Other processes where the photon chiral-odd DA contributes were considered earlier \cite{earlier}

The leading twist chiral-odd photon distribution 
amplitude $\phi_\gamma(u)$ reads \cite{Braun}
\begin{eqnarray}\label{def3:phi}
\langle 0 |\bar q(0) \sigma_{\alpha\beta} q(x) 
   | \gamma^{(\lambda)}(k)\rangle =  i \,e_q\, \chi\, \langle \bar q q \rangle
 \left( \epsilon^{(\lambda)}_\alpha k_\beta-  \epsilon^{(\lambda)}_\beta k_\alpha\right)  
 \int\limits_0^1 \!dz\, e^{-iz(kx)}\, \phi_\gamma(z)\,,  \nonumber
\label{phigamma}
\end{eqnarray}    
where the normalization is chosen as $\int dz\,\phi_\gamma(z) =1$, 
and $z$ stands for the momentum fraction carried by the quark. The product of the quark condensate and of the magnetic susceptibility of the QCD vacuum
$\chi\, \langle \bar q q \rangle$ has been estimated \cite{BK} with the help of the QCD sum rules techniques to be of the order of 50 MeV  and a lattice estimate has recently been performed \cite{Bui}. The  distribution amplitude 
$\phi_\gamma(z)$ has a QCD evolution which drives it to an asymptotic form $\phi^{as}_\gamma(z) = 6 z (1-z)$.
Its $z-$dependence at non asymptotic scales is very model-dependent \cite{Bro}.

To access the photon DA, we consider the following process ($s_T$ is the transverse polarization of the nucleon):
\begin{equation}
\label{process}
\gamma(k,\epsilon) N (r,s_T)\to l^-(p)  l^+(p') X\,, 
\end{equation}
with $q= p+p'$ in the kinematical region where $Q^2=q^2$ is large and the transverse component $ |\vec Q_\perp |$ 
of $q$ is of the same order as $Q$. 
Such a process  occurs either through a Bethe-Heitler amplitude  where the initial photon 
couples to a final lepton, or through Drell-Yan type amplitudes  where the final leptons originate from 
a virtual photon. 
Among these Drell-Yan processes, one must distinguish the cases 
where the real photon couples directly (through the QED coupling) to quarks or 
through its quark content.  
We  define the amplitude ${\cal A}_\phi$ as the  contribution where the photon couples to the strong 
interacting particles through its lowest twist-2 chiral odd distribution 
amplitude.  
One can easily see by inspection that interfering  the amplitude ${\cal A}_\phi$ with a pointlike amplitude, one gets at the level of twist 2 (and with vanishing quark masses) a contribution to  nucleon transverse spin dependent observables.
The cross section for reaction (\ref{process}) can  be decomposed as
\begin{eqnarray}\label{cs}
\frac {d\sigma}{d^4Q \,d\Omega}  =  \frac {d\sigma_{BH}}{d^4Q\,d\Omega} +  \frac {d\sigma_{DY}}{d^4Q\,d\Omega} +   \frac {d\sigma_{\phi}}{d^4Q\,d\Omega} +  \frac {\Sigma d\sigma_{int}}{d^4Q\,d\Omega}\,,\nonumber
\end{eqnarray}
where $\Sigma d\sigma_{int}$ contains various interference terms, while
the transversity dependent  differential cross section (we denote $\Delta_T \sigma = \sigma(s_T) - \sigma(-s_T)$) reads
\begin{eqnarray}\label{cst}
&&\frac {d\Delta_T \sigma}{d^4Q\,d\Omega}  = \frac {2d\sigma_{\phi int}}{d^4Q\,d\Omega}  \,, \nonumber
\end{eqnarray}
 where $d\sigma_{\phi int}$ contains only interferences between the amplitude ${\cal A}_\phi$ and the other amplitudes. Moreover, one may use the distinct charge conjugation property (with respect to the lepton part) of the Bethe Heitler amplitude to select the interference between ${\cal A}_\phi$ and the Bethe-Heitler amplitude :
\begin{eqnarray}\label{CAcst}
&&\frac {d\Delta_T \sigma (l^-) - d\Delta_T \sigma (l^+) }{d^4Q\,d\Omega}  = \frac {4d\sigma_{\phi BH}}{d^4Q\,d\Omega}  \,. \nonumber
\end{eqnarray}
We thus have a  differential cross section  proportional to  the photon distribution
 amplitude $\Phi_\gamma(z=\frac{\alpha Q^2}{ Q^2+\vec Q_\perp ^2})$ and to the nucleon transversity $h_1$. 

\section{Accessing transition distribution amplitudes}
Investigations of GPDs have been very important in the recent years since they are new QCD objects
 which carry much information on the hadronic structure. A further generalisation of the GPD
  concept has been proposed in cases where the initial and final states are 
different hadronic states \cite{FS}.
  If those new hadronic objects are defined through a quark-antiquark operator (meson to meson or meson to photon transition), we call them {\it mesonic}
 transition distribution amplitudes (TDA)~\cite{TDApigamma},
  if they are defined through a three quark operator (baryon to meson or baryon to photon transition), 
we call them {\it baryonic} transition distribution amplitudes~\cite{TDApiproton}. The TDAs can be accessed in various exclusive processes in $\gamma \gamma$  and  electron proton collisions and in proton-antiproton annihilation. 

Let us give some details for the most interesting case, namely the proton to pion TDAs which should measure the pion cloud around the hard valence quark core in the nucleon.
There are eight leading twist $p \to \pi^0$ TDAs but only three of them contribute in forward kinematics (i.e. at $\Delta_T=0$) and thus dominate; they are defined as $ V^{p\pi^0}_{i}\!\!(x_i,\xi, \Delta^2)$, 
$A^{p\pi^0}_{i}\!\!(x_i,\xi, \Delta^2)$ and 
$T^{p\pi^0}_{i}\!\!(x_i,\xi, \Delta^2)$  are defined  as
(${\cal F}$ denotes a Fourier transform), 
\begin{eqnarray}
\label{eq:TDApi0proton}
&&  {\cal F}\Big(\langle     \pi^0(p_\pi)|\, \epsilon^{ijk}u^{i}_{\alpha}(z_1 n) 
u^{j}_{\beta}(z_2 n)d^{k}_{\gamma}(z_3 n)
\,|P(p_p,s_p) \rangle \Big)=   
\frac{i}{4}\frac{f_N}{f_\pi}\Big[ V^{p\pi^0}_{1} (\ks p C)_{\alpha\beta}(u^+(p_p,s_p))_{\gamma} \nonumber\\
&&+A^{p\pi^0}_{1} (\ks p\gamma^5 C)_{\alpha\beta}(\gamma^5 u^+(p_p,s_p))_{\gamma} 
 +T^{p\pi^0}_{1} (\sigma_{p\mu} C)_{\alpha\beta}(\gamma^\mu u^+(p_p,s_p))_{\gamma}\Big]\;,\nonumber
\end{eqnarray}
where  $\sigma^{\mu\nu}= 1/2[\gamma^\mu, \gamma^\nu]$, $C$ is the charge 
conjugation matrix and 
 $u^+$ is the large component of the nucleon spinor.
 For these three TDAs, a soft meson theorem allows us to use the following expressions for   large $\xi$
\begin{eqnarray}
\label{eq:softp}
\{V^{p\pi^0}_1,A^{p\pi^0}_1,T^{p\pi^0}_1\}(x_1,x_2,x_3,\xi,0) = \frac{1}{4 \xi}  \{V^p, A^p,3 T^p\} (\frac{x_1}{2\xi},\frac{x_2}{2\xi},\frac{x_3}{2\xi}),\nonumber
\end{eqnarray}
where $V^{p}$, $A^{p}$ and $T^{p}$ are the proton DAs. How does one go beyond this large $\xi$ limit? What is the 
$t =\Delta^2-$dependence of these TDAs, which can be interpreted after a Fourier transform as an impact parameter picture of the pion cloud in the nucleon ? These questions have up to now no answers. Contrarily to the GPD case, there is no constraint for the $\xi \to 0$ limit. 
A few model calculations \cite{TDAmodel} have been performed but no lattice calculation has yet derived any value of the first $x_i-$ moments of these functions. We think that this new domain is very interesting and eagerly wait for progress in this  field. 

\section{Accessing partonic quantities through AdS}
Many talks in this workshop have been devoted to  recent advances in the proposal to use AdS tools to try to solve QCD in the strong regime. If this strategy were successful, non-perturbative QCD objects such as partonic distributions might be extracted from  scattering amplitudes calculated before any factorization procedure. To illustrate 
  how the partonic picture may emerge from a simple scenario based on
  the AdS/QCD correspondencee, we critically examined \cite{PRSW} the question of scaling of the Deep Inelastic
  Scattering (DIS) process in the medium Bjorken $x$ region on a scalar
  boson, following the strategy defined in \cite{PS}. Our aim was to enquire how one could recover the two main features of the partonic description of DIS, namely the facts that

- the amplitude scales as $Q^0$ (up to logarithmic modifications)
  at large $Q^2 $ and fixed Bjorken variable ($x_{Bj}= Q^2/s $),
  where $Q^2 = - q^2$ is the virtuality of the photon and $s$ the
  squared energy of the process, and this scaling behaviour is the
  signal that the electromagnetic current scatters on pointlike
  particles inside the hadron.

- the leading amplitude corresponds to the case of a
  transversally polarized virtual photon, which is the signal that
  these pointlike constituents are fermions. In other terms,  the longitudinal
  structure function $F_L(x)=F_2(x)-2xF_1(x)$ vanishes at leading order in $1/Q^2$. This is the Callan-Gross relation.

To perform the calculation, one has to sum over different type of hadrons in the final state. We choosed to allow scalar and vector mesons.  We demonstrated that the $Q^2$ dependence of the structure functions induced by vector intermediate
states is the same as for the scalar intermediate states. It is controlled only by the conformal dimension $\Delta_0$ of the scalar
initial state and does not depend on the conformal dimension
$\Delta_V$ of the vector intermediate states. Moreover, to get the right scaling behaviour, one needs to allow the conformal
  dimension of the hadronic initial field to equal $1$, which is at odds with the way it is fixed to get a reasonable description of electromagnetic form factors \cite{Brodsky}.  $\Delta_{0} = 1$ may be interpreted as the recognition
  that the incoming hadron fluctuates to an elementary field before
  scattering with the virtual photon. Most interestingly, it turns out
  that this value of $\Delta_{0}=1$ coincides with the unitarity bound
  on the dimension of scalar operator in four dimensions. 

With respect to the Callan-Gross relation, we observed that, to get the
  right polarization structure of the forward electroproduction  amplitude, one needs to add (at least) the scalar to
  scalar and scalar to vector hadronic amplitudes. However, this prevents the calculation from being predictive since an infinite number of parameters have to be fine-tuned. 

 This illustrates 
  how it is difficult to have the partonic picture  emerging from a simple scenario based on
  the AdS/QCD correspondence. 
We believe that this result is representative of a large class of non perturbative objects \cite{Hatta}

\section{Conclusion}
In conclusion, let us stress that, thanks to factorization properties of exclusive hard amplitudes in QCD, one can access and measure meaningful hadronic matrix elements describing the structure of the QCD vacuum and/or of hadrons. Experiments with intense beams at medium energies are spectacularly developping and one may hope to extract or at least severely constrain the normalization and functional forms of meson distribution amplitudes and proton to meson transition distribution amplitudes in the coming decade. However, one must admit that
non-perturbative QCD still seems in its infancy with respect to the theoretical computation of these quantities. 

\bigskip
This work is partly supported by the French-Polish scientific agreement Polonium and the Polish Grant N202 249235.

\end{document}